\documentstyle[12pt]{article}
\newlength{\extraspace}
\newlength{\extraspaces}
\newcommand{\be}{\begin{equation}
\addtolength{\abovedisplayskip}{\extraspaces}
\addtolength{\belowdisplayskip}{\extraspaces}
\addtolength{\abovedisplayshortskip}{\extraspace}
\addtolength{\belowdisplayshortskip}{\extraspace}}
\newcommand{\ee}{\end{equation}}

\newcommand{\ba}{\begin{eqnarray}
\addtolength{\abovedisplayskip}{\extraspaces}
\addtolength{\belowdisplayskip}{\extraspaces}
\addtolength{\abovedisplayshortskip}{\extraspace}
\addtolength{\belowdisplayshortskip}{\extraspace}}
\newcommand{\ea}{\end{eqnarray}}

\newcommand{\nonu}{\nonumber \\[.5mm]}
\newcommand{\A}{&\!\!\!}

\newcommand{\newsection}[1]{
\vspace{7mm} \pagebreak[3] \addtocounter{section}{1}
\setcounter{subsection}{0} \setcounter{footnote}{0}
\begin{center}
{\large {\bf \thesection. #1}}
\end{center}
\nopagebreak
\medskip
\nopagebreak \hspace{3mm}}

\begin{document}

\pagenumbering{arabic}

\begin{center}
{{\bf Reissner-Nordstr$\ddot{o}$m  Spacetime in the Tetrad Theory
of Gravitation}}
\end{center}
\centerline{ Gamal G.L. Nashed and Takeshi Shirafuji$^\ast$ }

\bigskip

\centerline{{\it Mathematics Department, Faculty of Science, Ain
Shams University, Cairo, Egypt }} \centerline{{ \it
$^{\ast}$Physics Department, Faculty of Science, Saitama
University, Saitama, Japan }}

\bigskip
 \centerline{ e-mail:nasshed@asunet.shams.edu.eg}

\hspace{2cm}
\\
\\
\\
\\
\\
\\
\\
\\

We give two classes of spherically symmetric exact  solutions  of
the couple gravitational and electromagnetic fields with charged
source in the tetrad theory of gravitation. The first solution
depends on an arbitrary function $H({R},t)$. The second solution
depends on a constant parameter $\eta$. These solutions reproduce
the same metric, i.e., the Reissner--Nordstr$\ddot{o}$m metric. If
the arbitrary function which characterizes the first solution and
the arbitrary constant of the second solution are set to be  zero,
then the two exact solutions will coincide  with each other. We
then calculate the energy content associated with these analytic
solutions using the superpotential method. In particular, we
examine whether these solutions meet  the condition which M\o ller
required for a consistent energy-momentum complex: Namely, we
check whether the total four-momentum of an isolated system
 behaves as a four-vector under Lorentz transformations. It is then found
that the arbitrary function should decrease faster than
$1/\sqrt{R}$ for $R\to \infty$. It is also shown that the second
exact solution meets the M\o ller's condition.

\newpage
\begin{center}
\newsection{\bf Introduction}
\end{center}

At present, teleparallel theory seems to be popular again, and
there is a trend of analyzing the basic solutions of general
relativity with teleparallel theory and comparing the results.  It
is considered as an essential part of generalized non-Riemannian
theories such as the Poincar$\acute{e}$ gauge theory \cite{Yi1}
$\sim$ \cite{BN} or metric-affine gravity \cite{HMM} as well as a
possible physical relevant geometry by itself-teleparallel
description of gravity \cite{HS1,NH}.  Teleparallel approach is
used for positive-gravitational-energy proof \cite{Me}. A relation
between spinor Lagrangian and teleparallel theory is established
\cite{TN}.  It has been shown that the teleparallel equivalent of
general relativity (TEGR) is not consistent in presence of
minimally coupled spinning matter \cite{Lm5}. Demonstration of the
consistency of the coupling of the Dirac fields to the TEGR has
been done \cite{ Me4}. However, it has been shown that this
demonstration is not correct \cite{OP,OP3}.

The tetrad theory  of gravitation  based on the geometry of
absolute parallelism \cite{PP}$\sim$\cite{AGP} can be considered
as the closest alternative to general relativity, and it has a
number of attractive features both from the geometrical and
physical viewpoints. Absolute parallelism is naturally formulated
by gauging spacetime translations and underlain by the
Weitzenb$\ddot{o}$ck spacetime, which is characterized by the
metric condition and by the vanishing of the curvature tensor.
Translations are closely related to the group of general
coordinate transformations which underlies general relativity.
Therefore, the energy-momentum tensor represents the matter source
in the field equation for the gravitational field just like in
general relativity.

{\it The tetrad formulation of gravitation was considered by M\o
ller in connection with attempts to define the energy of
gravitational field \cite{M6}$\sim$\cite{M7}. For a satisfactory
description of the total energy of an isolated system it is
necessary that the energy density of the gravitational field is
given in terms of first- and/or second-order derivatives of the
gravitational field variables. It is well-known that there exists
no covariant, nontrivial expression constructed out of the metric
tensor. However, covariant expressions that contain a quadratic
form of first-order derivatives of the tetrad field are feasible.
Thus it is legitimate to conjecture that the difficulties
regarding the problem of defining the gravitational
energy-momentum are related to the geometrical description of the
gravitational field rather than are an intrinsic drawback of the
theory \cite{Mj,MRTC}.}

M\o ller proposed \cite{M66} the three conditions which any
energy-momentum complex must satisfy:\\ (1) It must be an affine
tensor density which satisfies the conservation law.\\ (2) For an
isolated system the four-momentum is constant in time and
transform as a 4-vector under linear coordinate transformations.\\
(3) The superpotential transforms as a tensor density of rank 3
under the group of the spacetime transformations.\\ Then he showed
\cite{M7} that such an energy-momentum complex can be constructed
in the tetrad theory of gravitation.

It is the aim of the present work to find  spherically  symmetric
solutions in the tetrad theory of gravitation  for the coupled
gravitational and electromagnetic fields. We obtain two classes of
exact analytic solutions, and then calculate the energy of these
solutions using the superpotential given by M\o ller \cite{M7} and
Mikhail et.al. \cite{MWHL}. We shall then confirm that these
solutions meet the M\o ller's conditions when the asymptotic
conditions are imposed appropriately.
\newpage
The general form of the tetrad field, ${b_i}^\mu$, having
spherical symmetry was given by Robertson \cite{Rh}. In the
quasi-orthogonal coordinate system it can be written
as\footnote{In this paper Latin indices $(i,j,...)$ represent the
vector number, and Greek indices $(\mu,\nu,...)$ represent the
vector components. All indices run from 0 to 3. The spatial part
of Latin indices are denoted by $(a,b,...)$, while that of Greek
indices by $(\alpha, \beta,...).$
 }

\ba {{b_0}^0} \A= \A A, \quad {b_\alpha}^0 = C x^a, \quad
{b_0}^\alpha = D x^\alpha \nonu
{b_a}^\alpha \A= \A \delta_a^\alpha B + F x^a x^\alpha +
\epsilon_{a \alpha \beta} S x^\beta,
 \ea where {\it A}, {\it C},
{\it D}, {\it B}, {\it F}, and {\it S} are functions of ${\it t}$.
It can be shown that the functions $D$ and $F$ can be eliminated
by coordinate transformations \cite{HS7,SNH}, i.e., by making use
of freedom to redefine $t$ and $r$, leaving the tetrad field (1)
having four unknown functions in the quasi-orthogonal coordinates.
Thus the tetrad field (1) without the functions $D$ and $F$ will
be used in the following sections for the calculations of the
field equations of gravity and electromagnetism but in the
spherical polar coordinate.

In \S 2 we derive the field equations for the coupled
gravitational and electromagnetic fields. In \S 3 we first apply
the tetrad field (1) without the $S$-term to the derived field
equations.   We then give derivation for the general solution
without the $S$-term, and express the exact solution in terms of
an arbitrary function denoted by $H({R},t)$. A Relation between
this solution and a previous one \cite{Ng} is also established in
\S 3.  We also study the general, spherically symmetric solution
with a non-vanishing $S$-term in \S 3.
 In \S 4 we calculate the energy content of these two exact analytic solutions.
 Following M\o ller \cite{M66}, we require that the total four-momentum of an isolated
system be transformed as a four-vector under global, linear
coordinate transformations. Using Lorentz transformations we show
that the arbitrary function $H(R,t)$ should decrease faster than
$1/\sqrt{R}$ for $R\to \infty$.  We also examine the asymptotic
behavior of the solution with the non-vanishing $S$-term and we
find that its associated energy  is consistent with the M\o ller's
condition. The final section is devoted to discussion and
conclusion.

\newsection{The tetrad theory of gravitation and electromagnetism}

In the Weitzenb{\rm $\ddot{o}$}ck spacetime the fundamental field
variables describing gravity are a quadruplet of parallel vector
fields \cite{HS7} ${b_i}^\mu$, which we call the tetrad field in
this paper, characterized by \be D_{\nu} {b_i}^\mu=\partial_{\nu}
{b_i}^\mu+{\Gamma^\mu}_{\lambda \nu} {b_i}^\lambda=0, \ee where
${\Gamma^\mu}_{\lambda \nu}$ define the nonsymmetric affine
connection coefficients. The metric tensor $g_{\mu \nu}$ is given
by $g_{\mu \nu}= \eta_{i j} {b^i}_\mu {b^j}_\nu$ with the
Minkowski metric $\eta_{i j}=\textrm {diag}(-1,+1,+1,+1)$.
Equation (2) leads to the metric condition and the identically
vanishing curvature tensor.

 The gravitational Lagrangian $L_G$ is an invariant constructed
 from $g_{\mu \nu}$ and the contorsion tensor
  $\gamma_{\mu \nu \rho}$ given by \be \gamma_{\mu \nu \rho} =
{b^i}_{\mu}b_{i \nu; \ \rho} \,, \ee where the semicolon denotes
covariant differentiation with respect to Christoffel symbols. The
most general gravitational Lagrangian density invariant under
parity operation is given by the form \cite{HN,HS7}
 \be
{\cal L}_G  =  \sqrt{-g} L_G = \sqrt{-g} \left( \alpha_1 \Phi^\mu
\Phi_\mu
 + \alpha_2 \gamma^{\mu \nu
\rho} \gamma_{\mu \nu \rho}+ \alpha_3 \gamma^{\mu \nu \rho}
\gamma_{\rho \nu \mu} \right) \ee
 with $g = {\rm det}(g_{\mu
\nu})$ and $\Phi_\mu$ being the basic vector field defined by
 $\Phi_\mu = {\gamma^\rho}_{\mu \rho}$.  Here $\alpha_1,
\alpha_2,$ and $\alpha_3$ are constants determined
 such that the theory coincides with general relativity in the weak
 fields \cite{HN,M7}:
\be
 \alpha_1=-{1 \over \kappa}, \qquad \alpha_2={\lambda \over
\kappa}, \qquad \alpha_3={1 \over \kappa}(1-\lambda), \ee
 where
$\kappa$ is the Einstein constant and  $\lambda$ is a free
dimensionless parameter\footnote{Throughout this paper we use the
relativistic units, $c=G=1$ and
 $\kappa=8\pi$.}. The vanishing of this dimensionless parameter
 will reproduce the teleparallel equivalent theory of general
 relativity.

The electromagnetic Lagrangian  density ${\it L_{e.m.}}$ is
\cite{KT}  \be {\it L_{e.m.}}=-\displaystyle{1 \over 4} g^{\mu
\rho} g^{\nu \sigma} F_{\mu \nu} F_{\rho \sigma}, \ee with $F_{\mu
\nu}$ being given by\footnote{Heaviside-Lorentz rationalized units
will be used throughout this paper} $F_{\mu \nu}=
\partial_\mu A_\nu-\partial_\nu A_\mu$.

The gravitational and electromagnetic field equations for the
system described by ${\it L_G}+{\it L_{e.m.}}$ are the following:

 \be G_{\mu \nu} +H_{\mu \nu} =
-{\kappa} T_{\mu \nu}, \ee \be K_{\mu \nu}=0, \ee \be
\partial_\nu \left( \sqrt{-g} F^{\mu \nu} \right)=0 \ee
with $G_{\mu \nu}$ being the Einstein tensor of general
relativity.  Here
 $H_{\mu \nu}$ and $K_{\mu \nu}$ are defined by \be H_{\mu \nu}
= \lambda \left[ \gamma_{\rho \sigma \mu} {\gamma^{\rho
\sigma}}_\nu+\gamma_{\rho \sigma \mu} {\gamma_\nu}^{\rho
\sigma}+\gamma_{\rho \sigma \nu} {\gamma_\mu}^{\rho \sigma}+g_{\mu
\nu} \left( \gamma_{\rho \sigma \lambda} \gamma^{\lambda \sigma
\rho}-{1 \over 2} \gamma_{\rho \sigma \lambda} \gamma^{\rho \sigma
\lambda} \right) \right],
 \ee
and \be K_{\mu \nu} = \lambda \left[ \Phi_{\mu,\nu}-\Phi_{\nu,\mu}
-\Phi_\rho \left({\gamma^\rho}_{\mu \nu}-{\gamma^\rho}_{\nu \mu}
\right)+ {{\gamma_{\mu \nu}}^{\rho}}_{;\rho} \right], \ee and they
are symmetric and antisymmetric tensors, respectively. The
energy-momentum tensor $T^{\mu \nu}$ is given by  \be T^{\mu
\nu}=-g_{\rho \sigma}F^{\mu \rho}F^{\nu \sigma}+\displaystyle{1
\over 4} g^{\mu \nu} F^{\rho \sigma} F_{\rho \sigma} \ee

It can be shown \cite{HS7} that in spherically symmetric case the
antisymmetric part of the field equation (8) implies that the
axial-vector part of the torsion tensor, $a_\mu =
(1/3)\epsilon_{\mu\nu\rho\sigma}\gamma^{\nu\rho\sigma}$, should be
vanishing.  Then the $H_{\mu\nu}$ of (10) vanishes, and the field
equations (7)$\sim$(9) reduce to the coupled Einstein-Maxwell
equation in teleparallel equivalent of general relativity.  The
equation (7) then determines the tetrad field only up to local
Lorentz transformations
\[
{b^k}_\mu \to {\Lambda(x)^k}_{\ell}\, {b^{\ell}}_\mu\,,
\]
 which retain the condition $a_\mu =0$.  Hereafter we shall refer to this
 property of the field equations as {\it restricted local Lorentz
 invariance}.

\newsection{ Family of Reissner-Nordstr$\ddot{o}$m solutions}

 In this section we are going to study two cases of
the tetrad field(1).\\ {\underline{\it  Case I: The vanishing S-term.}}\\

 For the tetrad field (1) without the $S$-term the
axial-vector part of the torsion tensor, $a_{\mu}$, is identically
vanishing, and the remaining field equations possess the
restricted local Lorentz invariance. Thus, the general solution
for the tetrad field (1) without the $S$-term can be obtained from
the diagonal tetrad field for the Reissner-Nordstr$\ddot{o}$m
metric by a {\it local Lorentz transformation} which keeps
spherical symmetry \cite{SNH}
  \be
\left(\Lambda_{k l} \right) = \left( \matrix{ -L &  H \sin\theta
\cos\phi & H \sin\theta \sin\phi &  H \cos\theta \vspace{3mm} \cr
- H \sin\theta \cos\phi & 1+\left(L-1 \right)\sin^2\theta
\cos^2\phi &\left(L-1 \right)\sin^2\theta \sin\phi \cos\phi
&\left(L-1 \right)\sin\theta \cos\theta \cos\phi \vspace{3mm} \cr
- H \sin\theta \sin\phi &\left(L-1 \right) \sin^2\theta \sin\phi
\cos\phi &1+\left(L-1 \right)\sin^2\theta \sin^2\phi &\left(L-1
\right)\sin\theta \cos\theta \sin\phi \vspace{3mm} \cr - H
\cos\theta &\left(L-1 \right)\sin\theta \cos\theta \cos\phi
&\left(L-1 \right)\sin\theta \cos\theta \sin\phi  &1+\left(L-1
\right)\cos^2\theta \cr}\right), \ee
 where
$H$ is an arbitrary function of $t$ and $R$, and
\[L=\sqrt{H^2+1}.\]
 Namely, we see that
  \be {b_i}^\mu= \eta^{kl} \Lambda_{ik}\, {b_l}^{(\small 0)  \mu} \ee
 is the most general, spherically
symmetric solution without the {\it S}-term. Here ${b_l}^{ (\small
0) \mu}$ is the diagonal tetrad field which is given in the
spherical polar coordinates by \cite{Ng2}
 \be \left({b_l}^{ (\small 0) \mu} \right)=
\left( \matrix{ \displaystyle{1 \over {X}} &0 & 0 & 0 \vspace{3mm}
\cr 0 & {X} \sin\theta \cos\phi  & \displaystyle{\cos\theta
\cos\phi \over {R}}
 & -\displaystyle{ \sin\phi  \over {R} \sin\theta} \vspace{3mm} \cr
0 & {X} \sin\theta \sin\phi  & \displaystyle{\cos\theta \sin\phi
\over {R}}
 & \displaystyle{\cos\phi \over {R} \sin\theta} \vspace{3mm} \cr
0 & {X} \cos\theta & -\displaystyle{\sin\theta \over {R}} & 0 \cr
} \right), \ee
 where $X$ and $R$ are defined by
 \be X=\left[1-\displaystyle{2m \over R}+\displaystyle{q^2 \over
R^2} \right]^{1/2},
 \qquad \qquad  R={r/B}.\ee The explicit form of the ${b_i}^\mu$ is then given
by \be \left({b_i}^\mu \right) = \left( \matrix{ \displaystyle{ L
\over {X}} & H {X} &0 & 0 \vspace{3mm} \cr \displaystyle{H
\sin\theta \cos\phi \over {X} } & L {X} \sin\theta \cos\phi &
\displaystyle{ \cos\theta \cos\phi \over {R}} &
-\displaystyle{\sin\phi \over {R} \sin\theta} \vspace{3mm} \cr
\displaystyle{H \sin\theta \sin\phi \over {X}} & L {X} \sin\theta
\sin\phi & \displaystyle{ \cos\theta \sin\phi \over {R}} &
\displaystyle{ \cos\phi \over {R} \sin\theta} \vspace{3mm} \cr
\displaystyle{H \cos\theta \over {X}} & L {X} \cos\theta
&-\displaystyle{\sin\theta \over {R}} & 0 \cr}\right). \ee

If we apply the  tetrad field (17) to the field
 equations (7)$\sim$(9) then, the vector potential $A_\mu$, the antisymmetric electromagnetic
 tensor $F_{\mu \nu}$ and  ${T_{\mu}}^{\nu}$ take
 the form \be A_t(R)=-\displaystyle{q \over 2\sqrt{\pi} R}, \qquad
  F_{R t}=-\displaystyle{q \over 2\sqrt{\pi} R^2},\qquad
  {T_0}^0={T_1}^1=-{T_2}^2=-{T_3}^3=\displaystyle{q^2 \over 8 \pi
  R^4}.\ee
The metric associated with the tetrad field (17) is by definition
given by the  Reissner-Nordstr$\ddot{o}$m solution.

Now let us compare the solution (17) with that given before:
Nashed \cite{Ng} obtained a solution with an arbitrary function
${\cal B}$ for the tetrad (1) with three unknown function in the
spherical polar coordinate.  The tetrad field of that solution can
be obtained from (17) if the function $H$ is chosen as \be
H=\frac{\left[{R}^2 {\cal B}'^2-2{R} {\cal B}'+\displaystyle{2m
\over {R}}-\displaystyle{q^2 \over {R}^2} \right]^{1/2}}{X}.
 \ee
{\underline{\it  Case II: The non-vanishing S-term.}}\\

Let us next look for spherically symmetric solutions of the form
(1) with non-vanishing $S$-term by using the result that the
antisymmetric part of the field equation (8) requires the
axial-vector part of the torsion tensor, $a_{\mu}$, to be
vanishing for spherically symmetric case \cite{HS7}.  For this
purpose we start with the tetrad field (1) with the six unknown
functions of {\it t} and {\it r}. In order to study the condition
that the $a_{\mu}$ vanishes it is convenient to start from the
general expression for the covariant components of the tetrad
field ${b^i}_{\mu}$,
 \ba
{b^0}_{\scriptstyle{0}} \A= \A -\check{A}, \quad {b^a}_{
\scriptstyle{0}}= \check{C} x^a, \quad  {b^0}_{
\scriptstyle{\alpha}}= -\check{D} x^\alpha \nonu
{b^a}_{\scriptstyle{\alpha}} \A= \A \delta_{a \alpha}
\check{B}+\check{F} x^a x^\alpha + \epsilon_{a \alpha \beta}
\check{S} x^\beta, \ea
 where the six unknown functions,
$\check{A}$, $\check{C}$, $\check{D}$, $\check{B}$, $\check{F}$,
$\check{S}$, are connected with the six unknown functions of (1).
We can assume without loss of generality that the two functions,
$\check{D}$ and $\check{F}$, are vanishing by making use of the
freedom to redefine ${\it t}$ and ${\it r}$ \cite{HS7,SNH}. We
transform the tetrad field (20) to the spherical polar coordinates
 ($t,r,\theta,\phi$):
 \be \left(b_{i
\scriptstyle{\mu}} \right)= \left( \matrix{ \check{A} & 0 & 0 & 0
\vspace{3mm} \cr r \check{C} \sin\theta \cos\phi & \check{B}
\sin\theta \cos\phi & r \check{B}\cos\theta
\cos\phi+r^2\check{S}\sin\phi
 & -r \check{B} \sin\theta \sin\phi+r^2\check{S} \sin\theta \cos\theta \cos\phi \vspace{3mm} \cr
 r \check{C} \sin\theta \sin\phi & \check{B}
\sin\theta \sin\phi &r \check{B}\cos\theta
\sin\phi-r^2\check{S}\cos\phi  & r \check{B} \sin\theta
\cos\phi+r^2\check{S} \sin\theta \cos\theta \sin\phi \vspace{3mm}
\cr r \check{C} \cos\theta  & \check{B} \cos\theta & -r \check{B}
\sin\theta  & -r^2\check{S}\sin^2\theta \cr } \right). \ee
 The condition that the axial-vector part $a_{\mu}$ vanishes is
then expressed by \cite{SNH} \be
 0 = \sqrt{(-g)} a^\mu = \left\{
\matrix{ & 3 \check{B} \check{S}+ r(\check{B} \check{S}'
-\check{B}' \check{S}), \quad \mu=0, \hfill\cr &
 2 \check{C}\check{S}+ (\check{\dot{S}}\check{B} -\check{S}\check{\dot{B}}),
 \qquad \mu=1 \hfill\cr }\right.
 \ee
 with $\check{S}'= {d
\check{S}/dr}$  and $\check{\dot{S}}= {d \check{S}/dt}$. This
condition can be solved to give \be
 \check{C}=0, \quad \check{S}={{\eta} \over r^3} \check{B},
\ee where $\eta$ is a constant with dimension of
$\textrm{(length)}^2$.  The tetrad field (21) then gives the
following expression for the line element:
 \be
 ds^2=-\check{A}^2dt^2+\check{B}^2dr^2+r^2\check{B}^2\left(1+\displaystyle{\eta^2
 \over r^4}\right)d^2\Omega.
 \ee

The symmetric part of the field equations now coincides with the
Einstein equation. The metric tensor must be the
Reissner-Nordstr$\ddot{o}$m  solution when the Schwarzschild
 radial coordinate $R$ is used.  Therefore we choose the new radial
 coordinate
 \be R =r \check{B} \sqrt{1+{\eta^2 \over r^4}}\, , \ee
 and require that the line-element written in the coordinate $(t,R,\theta,\phi)$
 coincides with the Reissner-Nordstr$\ddot{o}$m metric.
Then we have
 \be
 \check{A}(r)=X(R)\,, \qquad \displaystyle{dR \over dr} = \check{B}(r)X(R)
 \ee
 where $X(R)$ is defined by (16) with the constants $m$ and $q$
 being interpreted as the total mass and the total charge,
 respectively, of the central body.  Eliminating $\check{B}$
 from (25) and the second equation of (26), we obtain a
 differential equation for $R(r)$, which can easily be solved to
 give \be
 r^2=|\eta|\sinh \,Y(R) \ee
 with the function $Y(R)$ being defined by
 \be
 Y(R)=2 \int \, \displaystyle{dR \over RX}=
 \ln \left[\displaystyle{\left(R-m+\sqrt{R^2-2mR+q^2}\right)^2 \over
 2|\eta|}\right]\, ,
 \ee
 where the additive integration constant is fixed in the last
 equation by requiring the asymptotic condition $r/R \to 1$ as
 $R \to \infty$.  Using (27) in (25) gives
 \be
 r\check{B}(r)=R \tanh Y(R)\,,
 \ee
 which together with (23) and (25) leads to
 \be
 r^2\check{S} = \displaystyle{\eta \over r^2}\left(r\check{B}\right)
 = \displaystyle{\eta \over |\eta|} \displaystyle{R \over \cosh\,Y(R)}\,.
 \ee

  Now it is straightforward to obtain the covariant components of
   the tetrad field, ${b^i}_{\mu}$,
with the non-vanishing $S$-term for the
Reissner-Nordstr$\ddot{o}$m solution in the coordinate system
$(t,R,\theta,\phi)$: The non-vanishing components are given by
 \ba
 {b^0}_{0} \A= \A  X \nonu
 {b^1}_{1} \A= \A
 \displaystyle{\sin\theta \cos\phi \over X} \nonu
{b^1}_{2} \A= \A R\left( \tanh Y \cos\theta \cos\phi +
 \displaystyle{\eta \over |\eta|} \displaystyle{\sin\phi \over \cosh Y} \right)
 \nonu
{b^1}_{3} \A= \A R\left( -\tanh Y \sin\phi + \displaystyle{\eta
\over |\eta|} \displaystyle{\cos\theta \cos\phi \over \cosh
Y}\right) \sin\theta
 \nonu
{b^2}_{1} \A= \A
 \displaystyle{\sin\theta \sin\phi \over X} \nonu
{b^2}_{2} \A= \A R\left( \tanh Y \cos\theta \sin\phi
-\displaystyle{\eta \over |\eta|} \displaystyle{\cos\phi \over
\cosh Y}\right)
 \nonu
{b^2}_{3} \A= \A R\left( \tanh Y \cos\phi +\displaystyle{\eta
\over |\eta|} \displaystyle{\cos\theta \sin\phi \over \cosh Y}
\right)\sin\theta \nonu
{b^3}_{1} \A= \A  \displaystyle{\cos\theta \over X} \nonu
{b^3}_{2} \A= \A -R \tanh Y \sin\theta
   \nonu
{b^3}_{3} \A= \A -R \left(\displaystyle{\eta \over |\eta|}
\displaystyle{\sin^2\theta
 \over \cosh Y}\right)\,.
 \ea
Or equivalently in the quasi-orthogonal coordinate system, in
which the spatial coordinates are given by $(x^\alpha) = (R
\sin\theta \cos\phi, R \sin\theta \sin\phi, R \cos\theta)$, the
space-space components ${b^a}_\alpha$ are expressed in a more
compact form:
 \be
 {b^a}_\alpha = \tanh Y\, \delta_{a\alpha} + \left(\displaystyle{1
 \over X}-\tanh Y\right)\displaystyle{x^ax^\alpha \over R^2}
 + \left(\displaystyle{\eta \over |\eta|} \displaystyle{1 \over
  \cosh Y}\right) \epsilon_{a\alpha\beta} \displaystyle{x^\beta
  \over R}\,.
 \ee
It is of interest to note that solution (31) is reduced to
solution (29) obtained before \cite{Ng2} when $q=0$ and $m$ is
replaced by $m(1-e^{-R^3/r1^3})$.

 Finally we notice that if the constant $\eta$ is set equal to zero the
tetrad field (31) reduces to the matrix inverse of the solution
(17) with $H=0$.

\newsection{The energy associated with each solution}

 The superpotential is given by \cite{M7, MWHL}
  \be {{\cal U}_\mu}^{\nu \lambda} ={(-g)^{1/2} \over
2 \kappa} {P_{\chi \rho \sigma}}^{\tau \nu \lambda}
\left[\Phi^\rho g^{\sigma \chi} g_{\mu \tau}
 -\lambda g_{\tau \mu} \gamma^{\chi \rho \sigma}
-(1-2 \lambda) g_{\tau \mu} \gamma^{\sigma \rho \chi}\right], \ee
where ${P_{\chi \rho \sigma}}^{\tau \nu \lambda}$ is \be {P_{\chi
\rho \sigma}}^{\tau \nu \lambda} \stackrel{\rm def.}{=}
{{\delta}_\chi}^\tau {g_{\rho \sigma}}^{\nu \lambda}+
{{\delta}_\rho}^\tau {g_{\sigma \chi}}^{\nu \lambda}-
{{\delta}_\sigma}^\tau {g_{\chi \rho}}^{\nu \lambda} \ee with
${g_{\rho \sigma}}^{\nu \lambda}$ being a tensor defined by \be
{g_{\rho \sigma}}^{\nu \lambda} \stackrel{\rm def.}{=}
{\delta_\rho}^\nu {\delta_\sigma}^\lambda- {\delta_\sigma}^\nu
{\delta_\rho}^\lambda. \ee
 The energy contained in the sphere with
radius $R$ is expressed by the surface integral \cite{M5} \be
E(R)=\int_{r=R} {{\cal U}_0}^{0 \alpha} n_\alpha d^2S\;, \ee where
$n_\alpha$ is the unit 3-vector normal to the surface element
$d^2S$.

Let us first discuss the solution given by (17). Calculating the
necessary components of the superpotential in the quasi-orthogonal
coordinates $(t,x^\alpha)$,
 \be {{\cal U}_0}^{0 \alpha}={2 {X}
x^\alpha \over \kappa {R^2}}\left(L-{X} \right),
 \ee
and substituting it into (36), we obtain \be E( {R})={X} {R}
\left(L-{X} \right)\,, \ee which depends on the arbitrary function
$H$.  Since this arbitrary function originates from the restricted
local Lorentz invariance of the field equations (7) and (9), the
result (38) shows that the energy content of a sphere with
constant $R$ is not invariant under restricted local Lorentz
transformations.

Next let us turn  to the solution (31). Calculating the necessary
components of the superpotential,
 \be
{{\cal U}_0}^{0 \alpha}={2X x^\alpha \over \kappa R^2}\left(\tanh
Y - X \right), \ee and substituting it into (36), we have
 \be
E(R)=X R (\tanh Y - X ).
 \ee
 For large $R$ this is rewritten as
 \be E(R)\cong m-\displaystyle{{q^2+m^2}\over {2R}}, \ee
  where only those terms up to order $O(1/R)$ are retained. In this
approximation the total energy is independent of the constant
$\eta$.  Finally we notice that the result (41) agrees with that
given before \cite{Ng,Ri}.

We now turn to study whether the obtained solutions (17) and (31)
satisfy the M\o ller's three conditions (1)$\sim$(3) recapitulated
in the Introduction.  Since the two conditions (1) and (3) are
satisfied in the tetrad theory of gravitation \cite{M7}, we shall
focus our attention on the condition (2).

We start with the solution (17). The asymptotic form of the tetrad
field ${b_i}^{\mu}$ is expressed up to $O(1/R^2)$ in the
quasi-orthogonal spatial coordinates $(x^\alpha)=(R\sin\theta
\cos\phi, R\sin\theta \sin\phi, R\cos\theta)$ by
 \ba {b_0}^0 \A= \A \left[1+\displaystyle {H^2 \over 2}+
 \displaystyle {m \over R}\left(1+\displaystyle {H^2 \over 2}\right)-\displaystyle
  {q^2 \over 2R^2}+\displaystyle {3m^2 \over
 2R^2}\right],
 \nonu
 {b_0}^{\alpha} \A= \A \left[H-\displaystyle{mH \over R } \right]
 n^\alpha,
 \nonu
 {b_a}^0 \A= \A \left[H+\displaystyle{mH \over R } \right]  n^a, \nonu
  {b_a}^{\alpha} \A= \A \delta_a^{\alpha}+\left[\displaystyle {H^2 \over 2}
  -\displaystyle {m \over R}\left(1+\displaystyle {H^2 \over 2}\right)+
  \displaystyle {q^2 \over 2R^2}-\displaystyle {m^2 \over 2R^2}\right]
   n^a n^{\alpha}.
 \ea
 We calculate the energy separately according to the
  asymptotic behavior of the arbitrary function $H(R)$.

\underline{Case I:} $H(R,t)\sim f(t)/\sqrt{R^{1-\epsilon}}$,
   where $0<\epsilon<1$.

 The calculation of energy for such asymptotic behavior shows
  that it is divergent as $R\rightarrow \infty$, so we exclude this
  case from our consideration.

\underline{Case II:} $H(R,t)\sim f(t)/\sqrt{R^{1+\epsilon}}$,
where $0<\epsilon$.

The calculation of energy for such an asymptotic behavior of
$H(R,t)$ gives \be E(R)=m-\displaystyle{{q^2+m^2} \over 2R},
 \ee
 up to order $O(1/R)$ in agreement with the result (41) for the solution
 (31), and the M\o ller's  condition (2) is satisfied.

 \underline{Case III:} $H(R,t)\sim f(t)/\sqrt{R}$.

 The non-vanishing components of the superpotential
 (33) are given asymptotically by
 \ba {{\cal U}_0}^{0 \alpha} \A =\A {2
n^\alpha \over \kappa R^2}\left[m-\displaystyle{q^2 \over
2R}+\displaystyle{f^2(t) \over 2}-\displaystyle{f^4(t) \over
8R}-\displaystyle{mf^2(t) \over 2R}-\displaystyle{m^2 \over
2R}\right],\nonu
{{\cal U}_\gamma}^{\beta 0} \A=\A {1 \over \kappa R^2}\left[\left(
\displaystyle { f^3(t) \over 4\sqrt{R}}+\displaystyle { m f(t)
\over \sqrt{R}} \right) {\delta_\gamma}^\beta-\left( \displaystyle
{ f^3(t) \over 4\sqrt{R}}-\displaystyle { m f(t) \over \sqrt{R}}
\right) n^\gamma n^\beta \right]=-{{\cal U}_\gamma}^{0 \beta
},\nonu
{{\cal U}_\gamma}^{\beta \alpha} \A=\A {1 \over \kappa R^2}
\left[\displaystyle{f^2(t) \over 2}-\displaystyle{f^4(t) \over
8R}-\displaystyle{m f^2(t) \over 2R}+\displaystyle{q^2 \over
2R}+\displaystyle{m^2 \over 2R}\right] \left(n^\alpha
{\delta_\gamma}^\beta-n^\beta {\delta_\gamma}^\alpha \right).
 \ea
 The energy-momentum complex ${\tau_\mu}^\nu$ is given by
 \be
 {\tau_\mu}^\nu={{{\cal U}_\mu}^{\nu \lambda}}_{, \ \lambda},\ee
and automatically satisfies the conservation law,
${{\tau_{\mu}}^\nu}_{,\ \nu} =0$.   The nonvanishing components of
${\tau_\mu}^\nu$ are expressed by
 \ba
 {\tau_0}^0 \A= \A {1 \over \kappa R^3 }\left[\displaystyle{q^2 \over
R}+\displaystyle{f^4(t) \over 4R}+\displaystyle{m f^2(t) \over
R}+\displaystyle{m^2 \over R}\right],\nonu
 {\tau_\alpha}^0  \A=\A {n^\alpha \over \kappa R^3} \left[ \displaystyle { f^3(t) \over
2\sqrt{R}}+\displaystyle { 3m f(t) \over \sqrt{R}} \right],\nonu
{\tau_{\alpha}}^{\beta}  \A=\A {1 \over \kappa R^3 }
\left[\displaystyle{3f^2(t) \over 2} n^{\alpha}
n^{\beta}-\left\{\displaystyle{f^2(t) \over
2}-\displaystyle{f^4(t) \over 4R}-\displaystyle{m f^2(t) \over
R}+\displaystyle{q^2 \over R}-\displaystyle{m^2 \over
R}\right\}{\delta_{\alpha}}^{\beta} \right]\,,
 \ea
 where we have neglected higher order terms of $1/R^4$.

Using (44) in (36) and keeping up to $O(1/R)$, we find that
 the energy $E(R)$ is given by
\be
 E(R)=m-\displaystyle{q^2 \over 2R}+\displaystyle{f^2(t) \over
2}-\displaystyle{f^4(t) \over 8R}-\displaystyle{m f^2(t) \over
2R}-\displaystyle{m^2 \over 2R},
 \ee
  where the first two terms
represent the standard value of the energy but there are extra
terms which contribute to the total energy.

 Now let us examine if condition (2) is satisfied or not in
the case III.   For this purpose we consider  the Lorentz
transformation  \be \bar{x}^0 = \gamma(x^0+vx^1), \qquad
\bar{x}^1=\gamma(x^1+vx^0), \qquad \bar{x}^2=x^2 \qquad
\bar{x}^3=x^3,\ee
 where the coordinates $\bar{x}^{\mu}$ represent
the rest frame of an observer moving with speed $v$ to the
negative direction of the $x^1$-axis, and $\gamma$ is given by
$\gamma=\displaystyle{1 \over \sqrt{1-v^2}}.$ Here the speed of
light  is taken to be unity. The energy-momentum in a volume
element $d^3\bar{x}$ on the hyperplane, $\bar{x}^0=\;$const., is
given by \cite{M5}
 \be {{\bar \tau}_\mu}^{\;\nu} d^3{\bar
x}=\displaystyle{\partial x^\rho \over
\partial {\bar x}^\mu} \displaystyle{\partial {\bar x}^\nu \over
\partial  x^\sigma}  {\tau_\rho}^\sigma
 \displaystyle{d^3x \over \gamma}.
 \ee
 Using equations (48) and (49),
  it is easy to calculate the
 components ${\bar{\tau}_\mu}^{\;0}\,$ as follows:
 \be
 {\bar{\tau}_\mu}^{\;0}
d^3{\bar x}=\displaystyle{\partial x^\rho \over
\partial {\bar x}^\mu} \left( {\tau_\rho}^0 +v{\tau_\rho}^1
\right ) d^3x. \ee

Integration of (50) over the three dimensional hyperplane with $
\bar{x}^0 =\;$constant gives
 \be  \int_{{\bar{x}^0}=
\textrm{constant}} {\tau_\mu}^0 d^3{\bar x} =
\displaystyle{\partial x^\rho \over
\partial {\bar x}^\mu}\left( \int_{x^0= \textrm{constant}}\left[{\tau_\rho}^0
+ v {\tau_\rho}^1\right] d^3x \right).
 \ee
  Using (44) and (45) allows us
to calculate the integral on the right-hand side of (51); for the
second term we have
 \be
  \int {\tau_\rho}^1 d^3x=\displaystyle{f^2
\over 6}\; \delta_{\rho}^1\,. \ee
 Thus, we obtain
 \be
{\bar P}_\mu = \displaystyle{\partial x^\rho \over
\partial {\bar x}^\mu}\left\{P_\rho+\displaystyle {v f^2 \over
6} {\delta_\rho}^1 \right\}, \ee
 or for the four components,
  \be
{\bar P}_\mu=\gamma \left\{ -\left(E+\displaystyle{v^2f^2 \over 6}
\right), v\left(E-\displaystyle{f^2 \over 6} \right),0,0 \right\},
\quad  where \quad E= \lim_{R \to \infty} E(R) =
m+\displaystyle{f^2 \over 2}, \ee by virtue of (47).
 Equation (54) shows that the four-momentum is not transformed
 as a 4-vector under Lorentz
transformations, and the M\o ller's condition (2) is not satisfied
in the case III! Therefore, this case of spherically symmetric
solutions, in which the components ${b_a}^0$ behave as
$1/\sqrt{R}$, is not physically acceptable although it gives
Reissner-Nordstr$\ddot{o}$m metric.

As for the solution with the non-vanishing $S$-term, the tetrad
field is given by (31) in the quasi-orthogonal coordinate system,
and for large $R$ it tends to the asymptotic form like\\
${b^i}_\mu = \delta^i_\mu + O(1/R)$, and therefore the M\o ller's
condition
 (2) is satisfied.

\newsection{Main results and discussion}

In this paper we have studied the coupled equations of the
gravitational and electromagnetic fields  in the tetrad theory of
gravitation, applying the most general spherically symmetric
tetrad field of the form (1) to the field equations. Exact
analytic solutions are obtained by studying two cases: The case
Without the S-term and the case with S-term.  In both cases we use
the previously derived result \cite{HS7} that the antisymmetric
part of the coupled field equations requires the axial-vector part
of the torsion tensor, $a_{\mu}$, to vanish.

We obtained two exact solutions in which the field equations
reduce to those of the Einstein-Maxwell theory in the teleparallel
equivalent of general relativity.  The metric is then that of the
Reissner-Nordstr$\ddot{o}$m solution. For the tetrad field of the
form (1) without the $S$-term, the condition $a_\mu =0$ is
automatically satisfied, and the most general solution can be
obtained from the diagonal tetrad field for the
Reissner-Nordstr$\ddot{o}$m metric by applying those local Lorentz
transformations which retain the form (1) without the $S$-term.
Since the general expression for those local Lorentz
transformations involves an arbitrary function denoted by
$H(R,t)$, the obtained solution (17) for the tetrad field also
involves this arbitrary function and reduces to the previous
solution \cite{Ng}  when the function $H$ is chosen appropriately
(19).

For the tetrad field of the form (1) with the non-vanishing
$S$-term, the solution (31) is derived by requiring the two
conditions: The one is $a_\mu=0$, and the other is that the metric
should coincide with the Reissner-Nordstr$\ddot{o}$m metric.  The
solution involves a constant parameter $\eta$. If this constant is
set equal to zero, the tetrad field (31) reduces to the
 matrix inverse of the solution (17) with $H=0$.

We have used the superpotential method \cite{M7, MWHL} to
calculate the energy of the isolated system described by the
obtained solutions, and studied the asymptotic conditions imposed
by the M\o ller's condition (2).

Concerning the solution (17), the energy $E(R)$, which is
contained within the sphere of radius $R$, is given by (38) and
depends on the arbitrary function. In other words, the energy
contained in a finite sphere does depend on the tetrad field we
use: This can be considered as a manifestation of the pseudotensor
character of the gravitational energy-momentum complex.

As for the asymptotic behavior of the function $H$, we conclude
that it must decrease faster than $1/\sqrt{R}$ for large $R$. In
this case the energy $E(R)$ takes the well-known form (43) for
large $R$, and the four-momentum is transformed as a 4-vector.
Thus all the M\o ller's condition are satisfied. We reach this
conclusion of the asymptotic behavior of the function $H$ in the
following manner.  If the arbitrary function $H$ decreases more
slowly than $1/\sqrt{R}$, the $E(R)$ will be divergent for $R\to
\infty$. If the arbitrary function $H$ behaves like $1/\sqrt{R}$
for large $R$, the associated energy does not agree with the
well-known one, and furthermore, as we have shown, the
four-momentum is not transformed as a 4-vector (54), violating the
M\o ller's condition (2).

 Next we have calculated the energy associated with  solution (31)
 with the non-vanishing $S$-term.  We obtain  expression (40)
 for  $E(R)$, which depends on the parameter $\eta$.  It follows
 from (40) that if $R\rightarrow 0$ then $E(R)\rightarrow \infty$,
and that  if $R\rightarrow \infty$ then $E(R)\rightarrow m$. It is
also shown  that the four-momentum behaves like a 4-vector,
indicating that this solution meets all the M\o ller's condition.
Thus we have obtained two exact solutions physically different
from each other as we have seen from the discussion of the energy.
They are identical only when the arbitrary function $H$ and the
arbitrary constant $\eta$ are set to be zero.

A summary of the main results is given in the table below. The
solutions of spherically symmetric  Reissner-Nordstr$\ddot{o}$m
 black hole are classified into two groups. The solution  without
 the $S$-term has an arbitrary function and the
 solution with the $S$-term has a constant parameter $\eta$.
\vspace{20mm}
\begin {center}
\centerline{Table I: Summary of the calculations of the}
\centerline {exact form of  energy of the solutions (34) and (47)}
\begin{tabular}{|c|c|c|c|} \hline
    & \multicolumn{2}  {|c|}  { Field equation}  & Energy
$E(R)$ \\   \cline{2-3}& & & \\  & Skew part & Symmetric part& \\ & & & \\ \hline & & & \\
Tetrad field & Satisfied &   Reissner- &${X} {R}\left( L -{X}
\right)$
 \\
without the $S$-term &  identically &  Nordstr$\ddot{o}$m
solution&

\\ & & & \\\hline & & &\\
Tetrad field & satisfied  &  Reissner- &$X R (\tanh Y -X )$
 \\
with $S$-term & when $a_\mu=0$ & Nordstr$\ddot{o}$m solution&  \\
& & & \\ \hline
\end{tabular}
\end {center}
\newpage
\vspace{20mm}
\begin {center}
 \centerline{Table II: Asymptotic behavior of the arbitrary
 function}
  \begin{tabular}{|c|c|c|} \hline & & \\
  Arbitrary Function & Energy $E(R)$ & Physically acceptable \\ & & \\ \hline
& & \\ $H\sim 1/ \sqrt{R^{1-\epsilon}}$&Divergent &No  \\ & &
\\ \hline & & \\ $H\sim
 1 / \sqrt{R^{1+\epsilon}}$&
$E(R)=m-\displaystyle{q^2+m^2 \over 2R}$& Yes\\ & & \\ \hline   &
&
\\  $H\sim
 1 / \sqrt{R}$& $ E(R)=m-\displaystyle{q^2+m^2 \over
2R}+\displaystyle{f^2(t) \over 2}-\displaystyle{f^4(t) \over
8R}-\displaystyle{mf^2(t) \over 2R}$& No\\ & & \\\hline
\end{tabular}
\end {center}

\vspace{2cm} \centerline{\large {\bf Acknowledgment}} One of the
authors (G.N.) would like to thank the Association of
International Education, Japan (AIEJ) for  follow-up research
scholarship  and also wishes to express his deep gratitude to all
members of Physics Department at Saitama University.


\newpage
\begin{thebibliography}{99}

\bibitem{Yi1} Y.Itin, {\it Class. Quant. Grav.} {\bf 19} (2002), 173

\bibitem{Yi} Y. Itin, {\it J. Math. Phys.} {\bf 46} (2005),
012501.

\bibitem{HNH}F.W. Hehl, Y. Neeman, J. Nitsch and P. Von der Heyde, {\it Phys.
Lett.} {\bf B78} (1978), 102.

\bibitem{H7} F.W. Hehl, {\it Four Lectures on Poincar$\acute{e}$ gauge theory},
{\rm in proceedings of the 6th. course on spin, Torsion and
supergravity}, held at Eric, Italy, (1979), eds., P.G. Bergmann,
V. De Sabbata (Plenum, N.Y.) P.5.

\bibitem{K9} T. Kawai, {\it phys. Rev. } {\bf D49}, (1994) 2862; {\it phys. Rev.
} {\bf D62} (2000), 104014.

\bibitem{BV} M. Blagojevi$\acute{c}$  and M. Vasili$\acute{c}$  {\it Class.
Quant. Grav.} {\bf 17} (2000), 3785.

\bibitem{BN} M. Blagojevi$\acute{c}$  and I. A. Nikoli$\acute{c}$ {\it Phys.
Rev. } {\bf D62} (2000), 024021.

\bibitem{HMM} F.W. Hehl, J.D. MacCrea, E.W. Mielke and Y. Ne'eman, {\it Phys. Rep.}
{\bf 258} (1995), 1.

\bibitem{HS1} K. Hayashi and  T. Shirafuji,  {\it Phys.\ Rev.\ }
{\bf D19} (1979), 3524.

\bibitem{NH} J.Nitsch and F.W. Hehl, {\it Phys. Lett.} {\bf B90} (1980), 98.

\bibitem{Me} E.W. Mielke, {\it Phys. Rev} {\bf D42} (1990), 3388.

\bibitem{TN} R.S. Tung and J.M. Nester, {\it Phys. Rev} {\bf D60} (1999),
021501.

\bibitem{Lm5} M.  Leclerc, {\it Phys. Rev} {\bf D71} (2005), 027503.

\bibitem{Me4} E.W. Mielke, {\it Phys. Rev} {\bf D69} (2004), 128501.

\bibitem{OP} Yu.N.Obukhov and J.G. Pereira, {\it Phys. Rev} {\bf D69} (2004), 128502.

\bibitem{OP3} Yu.N.Obukhov and J.G. Pereira, {\it Phys. Rev} {\bf D67} (2003), 044016.

\bibitem{PP}  C. Pellegrini and J. Plebanski, {\it Mat.\ Fys.\
 Scr.\ Dan.\ Vid.\ Selsk.\ }{\bf 2} (1963), no.3.

\bibitem{HN} K. Hayashi and  T. Nakano, {\it Prog.\ Theor.\ Phys.\ }{\bf 38}
(1967), 491.

\bibitem{HS7} K. Hayashi and  T.  Shirafuji, {\it Phys.\ Rev.\ }{\bf D19}
(1979), 3524.

\bibitem{Kw} W. Kopzy$\acute{n}$ski, {\it J.\ Phys.\ } {\bf A15} (1982),
493.

\bibitem{Nj} J.M. Nester, {\it Class.\ Quantum Grav.\ } {\bf 5} (1988),
1003.

\bibitem{KT} T. Kawai and N. Toma,  {\it Prog.\ Theor.\ Phys.\ } {\bf 87}
(1992), 583 .

\bibitem{AP} V.C. de Andrade and J.G. Pereira, {\it Phys.\ Rev.\ } {\bf D56}
(1997),4689.

\bibitem{AGP} V.C. de Andrade, L.C.T Guillen  and J.G. Pereira, {\it Phys.\
Rev.\ Lett.\ } {\bf 84} (2000), 4533; {\it Phys.\ Rev.\ } {\bf
D64} (2001), 027502.

\bibitem{M6}  C. M\o ller, {\it Mat.\ Fys.\ Medd.\ Dan.\ Vid.\ Selsk.\ }{\bf1} (1961),
 no.10\,;$\,$ {\it ``Tetrad fields and conservation laws in
general relativity"} in Proc. International School of Physics
``Enrico Fermi" ed. C. M\o ller, (Academic Press, London, 1962).

\bibitem{M66} C.  M\o ller,  {\it Mat.\ Fys.\ Medd.\ Dan.\ Vid.\ Selsk.\ }{\bf 35} (1966),
no.3.

\bibitem{M7} C.  M\o ller, {\it Mat.\ Fys.\ Medd.\ Dan.\ Vid.\ Selsk.\ } {\bf 39}  (1978),
 no.13.

\bibitem{Mj} J. W. Maluf, {\it J.\ Math.\ Phys.\ }{\bf 35} (1994), 335.

\bibitem{MRTC} J. W. Maluf, J. F. Da Rocha-neto, T. M. L. Toribio and K. H.
Castello-Branco,  {\it Phys.\ Rev.\ }{\bf D65} (2002), 124001.

\bibitem{MWHL} F.I. Mikhail,  M.I. Wanas, A. Hindawi and E.I. Lashin,  {\it
Int.\ J.\ Theor.\ Phys.\ }{\bf 32} (1993), 1627.

\bibitem{Rh} H.P. Robertson,  {\it Ann.\ of Math.\ (Princeton)} {\bf 33}
(1932), 496.


\bibitem{SNH} T. Shirafuji, G.G.L. Nashed, and K. Hayashi,   {\it Prog.\
Theor.\ Phys.\ } {\bf 95} (1996), 665.

\bibitem{Ng} G.G.L. Nashed, {\it Int. J. Mod. Phys. Lett. A}, {\bf 21} (2006), 3181.

\bibitem{Ng2}  G.G.L. Nashed,  {\it  Phys.\ Rev.\ } {\bf D 66} (2002),
064015.

\bibitem{MWLH} F.I. Mikhail,  M.I. Wanas,  E.I Lashin, and A. Hindawi,
 {\it Gen.\ Rel.\ Grav.\ }{\bf 26} (1994), 869.

\bibitem{M5}  C.  M\o ller, {\it Ann. of Phys. } {\bf 4} (1958), 347;  {\bf
12} (1961), 118.

\bibitem{Ri} I. Radinschi,  {\it Mod.\ Phys.\ Lett. }{\bf A16} (2001), 673;
 $\,$ gr-qc/0103110; $\,$ gr-qc/0104004.

\end{thebibliography}
\end{document}